\documentclass[11pt]{article} 

\usepackage[utf8]{inputenc} 
\usepackage{amssymb, amsmath}
\usepackage[normalem]{ulem}
\usepackage{footmisc}
\usepackage{abstract}
\usepackage{multicol}
\usepackage{url}
\usepackage{multirow}
\usepackage[none]{hyphenat}
\usepackage{times}

\usepackage{geometry} 
\usepackage{graphicx} 
\usepackage[small,bf]{caption}

\usepackage{booktabs} 
\usepackage{array} 
\usepackage{verbatim} 

\makeatletter
\newenvironment{tablehere}
  {\def\@captype{table}}
  {}

\newenvironment{figurehere}
  {\def\@captype{figure}}
  {}
\makeatother

\title{A new method for the determination of action integrals\\in the study of galactic dynamics}
\author{M. F. J. Fox, MPhys Thesis.\\ Merton College, Oxford. \\
Supervisor: Prof. James Binney. \\
Project Number: TP14. \\
Word Count: 7,307.}
\date{April 20, 2012} 

\begin{document}
\maketitle
\begin{abstract}
\noindent Action-angle coordinates are an essential tool for understanding the properties of the six dimensional phase space involved in orbits of stars in galactic potentials. A new method, which does not require specific knowledge of a generating function, is described, implemented and tested that calculates the actions of an orbit in an arbitrary potential of an integrable Hamiltonian given a set of Cartesian phase space points. The method chooses between the simple harmonic oscillator and isochrone potentials to fit the data using a Levenberg-Marquardt routine. An average is taken over the angle coordinates by calculating volumes in phase space using the metric free FiEstAS algorithm. The perfect ellipsoidal potential, with actions  chosen \emph{a priori}, is used to test the output of the algorithm, giving some results that agree within 1\%. Minimisation of a sampling error is discussed along with an identification of a source of noise in the data.
\end{abstract}

\begin{multicols}{2}

\section{Introduction}
The problem of the behaviour of stars in a galaxy is of interest because it gives insight into the formation processes and evolution of these galaxies. Most notably it has led to the proposition of \emph{dark matter} in order to explain the observed velocities of stars \cite{darkmatter}. One can model a galaxy in a brute force sense by simply producing a simulation of the \(N\) stars (or dark matter particles), give them initial conditions and let them all interact with one another via the gravitational force. As well as being computationally time consuming, as typically \(N\approx10^{11}\) \cite{BT}, the insight gained from such methods is limited. In order to gain an understanding of the system it is the statistical distribution of orbits that are of interest (\textsection 4.1 of \cite{BT}) and these can be more easily reached through appropriate simplifications.


If one concentrates on a single star and releases it in a galaxy, one can ask, ``What trajectory will it follow?". As we do not want to do the integration of the \(N\)-body interactions, the orbit is calculated by smoothing the mass distribution of all the other stars so that we can produce a \emph{galactic potential} in which the star travels \cite{Dendy}. From the observed structures of galaxies the luminous mass distribution can be deduced and many different analytic forms of potential have been proposed (see chapter 2 of \cite{BT} for a review). 

The smoothing is valid because the gravitational force is purely attractive so no shielding of the force occurs. Thus, in a galaxy, the force on a star is dominated by the contribution from the large number of distant stars rather than that of the nearest neighbour stars (\textsection 1.2 of \cite{BT}). Further assumptions in this method include that the galaxy is in a steady state, such that the lifetime of the stars are much greater than one orbital period (table 11.1 of \cite{Dendy}). It is also implicit that no two stars in the model will collide, which is valid as the number density of stars in a typical galaxy is relatively low\footnote{Assuming a uniform distribution of stars in the Milky Way, with \(N=10^{11}\), radius of the disc \(10 kpc\) and thickness of \(0.5 kpc\) gives a number density of 0.6 stars per cubic parsec. Data from \cite{BT}.}. For a given potential, under these assumptions, the trajectory only depends on the initial conditions, which correspond to six constants of motion (\textsection 3.1.1 of \cite{BT}). By varying the initial conditions one can explore the orbits allowed by that potential.

The galactic potentials that will be considered here will always have at least three integrals of motion associated with them, which are constants that are not explicit functions of time (\textsection 3.1.1 of \cite{BT}). As an example of the simplest case, take a spherical potential: here we have the Hamiltonian and the three components of angular momentum being conserved. In the general case the Hamiltonian is conserved, along with at least two other quantities that are not necessarily the angular momenta. It is possible, for periodic systems, to use functions of these integrals as momentum coordinates \cite{Arnold}. One can do so provided the functions are isolating integrals and the conjugate coordinates to the momenta form a global coordinate system (see \textsection\ref{sec:AAcoords}). In this case the momenta are called actions and have conjugate coordinates called angles, which can be normalised with an amplitude of \(2\pi\).

The advantage of transforming to action-angle coordinates is that the equations of motion become exceedingly simple. The Hamiltonian is a function of the actions only and the angles increase linearly in time with constant frequencies \(\Omega_i = \dot\theta_i = \partial H/\partial J_i\) \cite{Schaums}. Furthermore the actions are adiabatic invariants, so do not change when the system is varied slowly, and thus become a useful tool in perturbation theory to treat non-steady state systems \cite{Arnold}\cite{BT}. For these reasons use of action-angle coordinates extends far beyond that in the study of galactic and planetary dynamics to include atomic, molecular, plasma and high energy physics \cite{Reiman}.


The three actions uniquely describe an orbit in a galactic potential, by labelling the phase space volume occupied by the angles \cite{Merritt}. The angles describe the location on a given orbit, but carry no defining information. It is then possible to reduce the six dimensional phase space to a three dimensional one, given just by the actions, which can then be drawn \cite{deZeeuw}. The locations of points within this phase space diagram allow a classification of the types of orbits allowed in a given potential by construction of the \emph{distribution function}. The distribution function is the fundamental description of the system and is used extensively in \(N\)-body simulations (\textsection 4.7.1 of \cite{BT}). 

Traditionally the action-angle coordinates of an arbitrary potential have been found by calculating the terms in an expansion of a generating function using best-fit methods \cite{Kaas}\cite{McGill}\cite{McMillan}\cite{Merritt}. Where the generating function maps the analytically known action-angle variables of a ``toy" Hamiltonian to the action-angle variables of the ``target" Hamiltonian.

This report details a new method for calculating the actions in an arbitrary galactic potential. The actions \(J_i(\vec{x},\vec{v})\) are found as functions of  the Cartesian phase space data points of an integrated orbit in the potential. Furthermore the method does not require explicit knowledge of the generating function and by using the FiEstAS algorithm \cite{Ascasiber} the required averaging is computationally fast. It is possible that the orbit could be one extracted from an \(N\)-body simulation and thus the method can be used to construct the distribution function and classify orbits in such simulations. 

The report develops as follows: In \textsection\ref{sec:background} the background theory on generating functions and action-angle coordinates is covered. Then \textsection\ref{sec:analyticpots} discusses the forms of potentials which have analytic expressions for the action-angle coordinates. \textsection\ref{sec:algtheory} covers the theory that justifies the approach taken for the algorithm. This is followed in \textsection\ref{sec:algpract} with an account of the practical implementation of the algorithm. \textsection\ref{sec:orbInt} covers details of the orbit integrator used to generate the required data to test the algorithm. In \textsection\ref{sec:ellipsoid} the ellipsoidal potential is introduced with its use in Cartesian coordinates outlined. \textsection\ref{sec:results} presents the results of using the ellipsoidal potential to test the algorithm, given known actions. A discussion of the results follows in \textsection\ref{sec:discussion}, detailing how the procedure had to be improved. Finally, in \textsection\ref{sec:conclusions} the report is summed up and possible future work discussed.

\section{Background theory}\label{sec:background}
First the concept of a generating function, which will be used extensively, is detailed. Then the theory behind action-angle coordinates and a description of how one finds their form for a given system is presented. For a detailed background in the mathematics that underpins this sections see \emph{Nash \& Sen 1983} \cite{Nash} for topology and \emph{Arnold 1989} \cite{Arnold} for classical mechanics.

\subsection{Generating functions}

Given two canonical coordinate systems \((\vec{q}, \vec{p})\) and \((\vec{Q}, \vec{P})\) the generating function \(S\) is a function of two of the variables, one from each of the two coordinate systems, that transforms between the two systems \cite{Schaums}. For example, if \(S = S(\vec{q}, \vec{P})\) then the other two coordinates are found from 
\begin{equation}\label{genfunc}
p_i = \frac{\partial S(\vec{q}, \vec{P})}{\partial q_i},\quad Q_i = \frac{\partial S(\vec{q}, \vec{P})}{\partial P_i}.
\end{equation}

\subsection{Action-angle coordinates}\label{sec:AAcoords} 

\begin{figurehere}
\includegraphics[width=\linewidth]{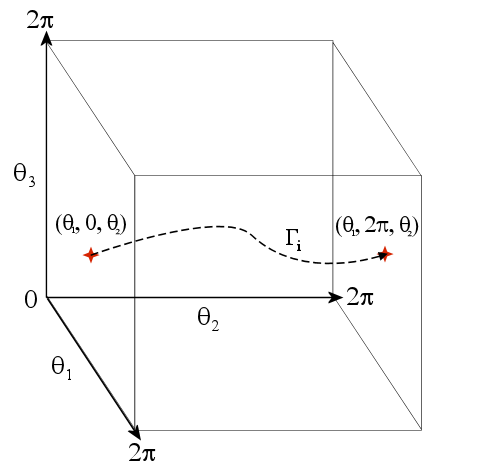}
\caption{\emph{Construction of a 3-torus, where the points on each face of the box are identified with those on the opposite face.}}
\label{fig:torus}
\end{figurehere}

In action-angle coordinates the trajectories of particles are described by the angles, whilst the constant actions label the orbit. The set of phase space coordinates on which \(J=constant\) is a \(n\)-torus, with \(n\) being the number of degrees of freedom of the system.  In the case in hand we have a 3-torus, so we have a cube with volume \((2\pi)^3\). The axes are labelled between \(0\) to \(2\pi\) and the coordinates at the planes on each axis at \(0\) and \(2\pi\) identified with each other, see figure \ref{fig:torus}. This description follows from Liouville's theorem \cite{Arnold}.


The actions themselves are isolating integrals, \(I(\vec{x},\vec{p})\). Isolating integrals are functions of the six phase space coordinates that are constant on smooth five dimensional surfaces in phase space \cite{BinneyWinter}, 
\begin{equation}\label{isolatingint}
I(\vec{x}(t), \vec{p}(t)) = \mathrm{constant}.
\end{equation}

As we assume that there exist three isolating integrals, the orbit we consider has the property of quasiperiodicity \cite{Arnold}\cite{BinneyWinter}. A quasiperiodic orbit is one in which the equation describing the temporal evolution of the coordinates of a particle in the potential, \(x(t)\), can be written as a discrete Fourier series, where the frequencies are integer linear combinations of three fundamental frequencies. The quasiperiodic nature of the system then directly relates to the angle coordinates as the fundamental frequencies are associated with the \(\Omega_i\) of the angles. Although not all potentials allow quasiperiodic orbits, so long as the orbits are close to being quasiperiodic the assumption of three isolating integrals should still hold \cite{Dendy}.

In order for the isolating integrals to be actions they must also have conjugate coordinates which form a global coordinate system \cite{BinneyWinter}. To form a global coordinate system the angles must describe trajectories that each take one entire loop around the torus, and they must return to the same point after the loop. The action is then defined \cite{BinneyWinter} as
\begin{equation}\label{actionInt}
J_i = \frac{1}{2\pi} \oint_{\Gamma_i} \vec{p}\cdot\mathrm{d}\vec{q},
\end{equation}
where the integral is along a closed path \(\Gamma_i\) that goes from a point on one face of the cube in figure \ref{fig:torus} to the associated point on the opposite face. It is important that a complete loop is made because otherwise the angles will only be local variables.



\subsection{Converting to action-angle coordinates}
When the data does not form a closed loop it is not possible to use the definition (\ref{actionInt}). But one can still transform to action-angle variables from a given canonical coordinate system by use of a generating function \cite{Arnold}. The problem is to find the appropriate generating function for the transformation. This can be done by solving the Hamilton-Jacobi equation, which in general is problematic, as it is non-linear. 
We look for a generating function \(S(\vec{x}, \vec{J})\) so that
\begin{equation}\label{ptheta}
p_i = \frac{\partial{S}}{\partial{x_i}}, \quad \theta_i = \frac{\partial{S}}{\partial{J_i}}.
\end{equation}
Then the time independent Hamilton-Jacobi equation takes the form \cite{Schaums}
\begin{equation}\label{HJ}
H\left(x_i, \frac{\partial S}{\partial x_i} \right) = E,
\end{equation}
where \(E\) is the energy and \(H\) is the Hamiltonian of the system. The method used to solve (\ref{HJ}) then usually relies on assuming a solution which is separable in each position variable \(x_i\) and thus the solution is reached through the normal separation of variables procedure \cite{Schaums}.

\section{Potentials with analytic expressions for the action-angle coordinates}\label{sec:analyticpots}
\begin{center}
\begin{figurehere}
\includegraphics[width=\linewidth]{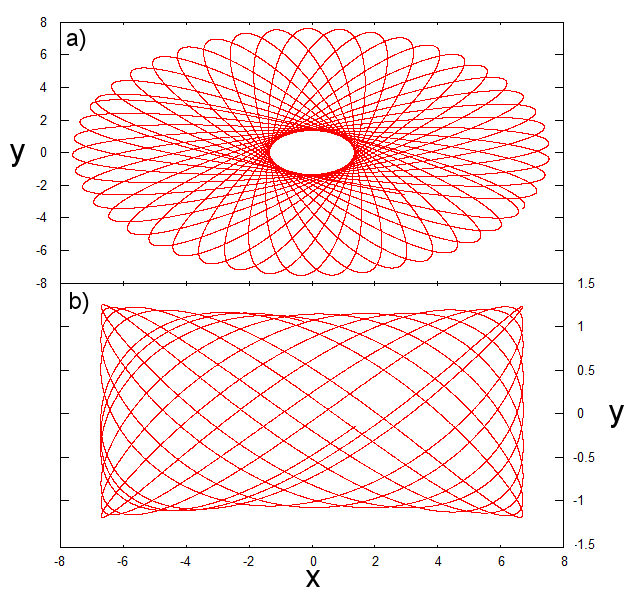}
\caption{\emph{Projection on to the x-y plane of (a) a loop orbit and (b) a box orbit. Arbitrary units of length.}}
\label{fig:boxloop}
\end{figurehere} 
\end{center}
There are a limited number of potentials for which the action-angle coordinates are analytically known in terms of the Cartesian coordinates. The algorithm, that will be described in \textsection\ref{sec:algtheory}, utilises these as ``toy" potentials to assign values of actions and angles to points from the integrated orbit in Cartesian space. 

The two potentials that are used are the isochrone and the simple harmonic oscillator (SHO) potentials \cite{BT}. These are used because the general form of trajectories in these potentials covers the two main types of bound orbits that can exist in an arbitrary galactic potential. The isochrone is an axisymmetric potential and produces loop orbits (figure \ref{fig:boxloop}a) which conserve angular momentum along a given axis. The SHO does not conserve angular momentum and produces box orbits (figure \ref{fig:boxloop}b). The choice of which type of potential to use is based on the type of orbit being fitted, which is discussed in \textsection\ref{sec:algpract}. 

\subsection{The potentials}
The isochrone potential (\textsection 2.2.2 of \cite{BT}) is
\begin{equation}
V(r) = \frac{-GM}{b + \sqrt{b^2 + r^2}},
\end{equation}
which in the limit \(r \gg b\) reduces to the Kepler potential and in the limit \(r \ll b\)  reduces to the spherical harmonic potential. \(GM\) and \(b\) are free parameters, that physically would represent the total mass, \(M\), of the galaxy and the characteristic radius, \(b\), at which the potential starts to fall as \(r^{-1}\) after being spherically harmonic (goes as \(r^2\)) in the centre. If these orbits are not closed then in position space they fill the volume of a 2-torus.

The SHO potential (\textsection 3.5.1 of \cite{BT})
\begin{equation}
V(\vec{x}) = \frac{1}{2} \sum_{i=1}^3 \omega_i^2 x_i^2
\end{equation}
gives box orbits, so named because if left to run over a sufficiently long time such an orbit fills a cuboid in position space. The \(\omega_i\) are free parameters that can be considered physically as the frequencies of oscillations in each coordinate direction.

\subsection{Conversion between Cartesian and action-angle coordinates}\label{sec:conversion}
The relationships between the two coordinate systems for the isochrone potential can be found in appendix \ref{sec:AAana}. The derivation for the SHO potential is outlined here because it is relatively simple and thus illustrates the method well (the two dimensional case is treated in \textsection 3.5.1 of \cite{BT} and is easily generalised to three dimensions). 

The Hamiltonian for the SHO is clearly separable into the three Cartesian components. Thus, the Hamilton-Jacobi equation is solved by writing the generating function as \(S = S_x(x, \vec{J}) + S_y(y, \vec{J}) + S_z(z, \vec{J})\) which then gives an expression that can be solved by separation of variables for \(S\) in terms of the \(x_i\). The actions are then found by considering the change in the generating function given by taking a loop around the torus of \(2\pi\) along one of the axes and then normalising by \(2\pi\). This gives
\begin{equation}\label{SHOJ}
J_i = \frac{p_i^2 + \omega_i^2 x_i^2}{2\omega_i}.
\end{equation}
Once the generating function is known the angles can be computed by taking the partial derivative given in the second equation of (\ref{ptheta})  so that
\begin{equation}\label{SHOtheta}
\theta_i =\mathrm{arctan}\Big(\frac{p_i}{\omega_i x_i}\Big).
\end{equation}

\section{Algorithm theory}\label{sec:algtheory}
\subsection{Calculating actions for arbitrary potentials}

The procedure for the algorithm is as follows: a set of phase space data points describing an orbit in an arbitrary potential have an analytic potential fitted to them based on the type of orbit. If the orbit is a loop, an isochrone potential is fitted and if it is a box then a harmonic potential is fitted. The fitted potential then plays the role of the ``toy" potential. Values of actions and angles for each of the data points can be assigned from the known relations between Cartesian coordinates and action-angle coordinates of these potentials. However, these actions will not be constants, because the potential that the orbit has been integrated in will differ from the isochrone or SHO potential. The actions of the ``target" potential are then extracted by considering a canonical transformation, where the key process involves an average being taken over the angle coordinates.

In order to perform the transformation a  generating function of the form \(S(\vec{\theta}, \vec{J'})\) is considered. This generating function transforms from the coordinates of the ``toy" potential \((\vec{\theta}, \vec{J})\) to the coordinates of the ``target" potential \((\vec{\theta'}, \vec{J'})\). Because the system is periodic we can expand the generating function as a Fourier series \cite{McGill} giving the general form 
\begin{equation}
S(\vec{\theta}, \vec{J'}) = \vec{\theta}\cdot\vec{J'} - i\sum_{\vec{n}\neq 0} S_{\vec{n}}(\vec{J'})\mathrm{exp}(i\vec{n}\cdot\vec{\theta}). 
\end{equation}
Remembering that \(\vec{J'}\) is a constant; then as the coefficients \(S_{\vec{n}}(\vec{J'})\) are functions of the \(\vec{J'}\) only, they too are constants. The \(\vec{n}\) are three-vectors with integer values. The \(\vec{J}\) are  computed through
\begin{equation}
J_i = \frac{\partial S(\vec{\theta}, \vec{J'})}{\partial \theta_i}
\end{equation}
which gives, on rearranging for the target action,
\begin{equation}\label{J'}
J'_i = J_i - \sum_{\vec{n}\neq 0}n_i S_{\vec{n}}(\vec{J'})\mathrm{exp}(i\vec{n}\cdot\vec{\theta}).
\end{equation} 
Now, on taking the average over the \(\theta\) variables
\begin{equation}
\vec{J'} = \langle\vec{J}\rangle,
\end{equation}
because the second term in (\ref{J'}) is a sum of periodic functions that on averaging over a whole period vanish. So all that needs to be done is to calculate
\begin{equation}\label{Jintegral}
\langle\vec{J}\rangle = \frac{1}{(2\pi)^3}\int_{\vec{\theta}}\vec{J}\mathrm{d}^3\theta.
\end{equation}

\section{Explaining the algorithm}\label{sec:algpract}
The code that has been written takes as an input a list of Cartesian phase space coordinates that describe an orbit in an arbitrary potential and gives an output of three actions for that orbit. Appendix \ref{sec:FlowDiagram} includes a basic flow diagram of the algorithm.

The algorithm first decides whether the orbit is best described as a loop or a box orbit. The choice between the two is based on the fact that in loop orbits there is a well defined sense of rotation about one of the axes, whereas this is not true for box orbits. Two methods of identifying this rotation have been tested. 

The first method computes the mean angular momentum along the three axes and the standard deviation of each mean. If the standard deviation is greater than the mean in all three directions the orbit is a box. Otherwise the angular momentum is conserved in at least one direction, giving a loop orbit.

In the second method the sign of the angular momentum is calculated at each point. If the sign changes then this clearly implies that there is no well defined sense of rotation about that axis. If the sign of all three components does change then one concludes that it is a box orbit. Otherwise it is a loop.

The second method has proven, during running the algorithm, to be more robust in identifying the two orbit types and has thus been chosen as the preferred method. This is because the first method requires a large number of data points fairly spread across the phase space of the orbit in order to accurately calculate the statistical quantities required for the analysis, whereas the second method can stop as soon as the signs on all three axes have changed.

In the case of a loop orbit it is important to identify which axis has the well defined sense of rotation. This is because that axis must be passed as the z-axis to the code which converts the Cartesian coordinates to action-angle coordinates.

The chosen toy potential (see \textsection \ref{sec:analyticpots}) is then fitted to the data points using a Levenberg-Marquardt fitting algorithm, which was written following \cite{LevMar}. This is done through the free parameters available in the potentials, plus an extra additive energy term, that adds another degree of freedom and allows for a better fit. Knowing the parameters it is then possible to assign the action and angle coordinates for each data point.  The Levenberg-Marquardt algorithm works by minimising \(\chi^2 = \sum_i(H_i-\langle H\rangle)^2\) through a damped Gauss-Newton algorithm \cite{TUDLevMar}. The \(H_i\) are the Hamiltonian of the chosen toy potential evaluated at each data point. 

The average of the actions over the angles is then taken following \textsection\ref{sec:algtheory}.  However, because we have a discrete set of data the integral in (\ref{Jintegral}) turns into a sum
\begin{equation}\label{discreteAverage}
J'_i = \langle J_i \rangle =  \frac{\sum_{\vec{\theta}} J_i \times \mathrm{Vol}(\vec{\theta})}{\sum_{\vec{\theta}}\mathrm{Vol}(\vec{\theta})}.
\end{equation}
The denominator is ideally \((2\pi)^3\) by the normalisation of the \(\theta\) coordinates, but remains an explicit sum in the algorithm so that the result is always properly normalised.

The volume is found by using the binary tree method of the FiEstAS algorithm \cite{Ascasiber}. This calculates the density of the data points in angle space and can easily be modified to give the volumes. FiEstAS works by systematically dividing the volume of the angle space into two, alternately along each coordinate axis, until only one point exists in each box. The size of the box then gives a measure of the volume around each angle space point. The method calculates the mean coordinate value of all the data points within a given box and then divides that box halfway between the two points nearest the mean, so that there are approximately an equal number of points on each side of the division.

The FiEstAS algorithm is the key to being able to compute the average in a short period of time, as previous methods could take hours, if not days, to calculate the volumes, whilst FiEstAS takes seconds \cite{Ascasiber}.

\section{Orbit Integrator}\label{sec:orbInt}

The orbit integrator follows the Runge-Kutta (RK) fifth-order method described in \emph{Numerical Recipes} \cite{Press}. Runge-Kutta methods are an extension of the simple Euler method of iteration where the next coordinate point \(x_{i+1}\) can be found from the previous point \(x_i\) as
\begin{equation}
x_{i+1} = x_i + v_i t,
\end{equation}
where \(v_i\) is the time derivative of \(x_i\) and \(t\) is the size of the time step. The time step has to be sufficiently small that the linear approximation holds. RK methods split the time step and evaluate the velocity at each intermediate point, then combine each term as a linear sum with coefficients selected to get a better estimate of the next position. The coefficients are usually based on the Taylor expansion of the position. In the fifth-order method used, each time step involves six separate function evaluations.

The RK method solves first order differential equations. We have Hamilton's equations \cite{Arnold} of the form
\begin{eqnarray}
\dot v_i &=&  - \frac{\partial{V}}{\partial{x_i}},\\
\dot x_i &=&  v_i,
\end{eqnarray}
which need to be solved simultaneously. The RK code was modified to evaluate both equations, at each step, through the integration.

The integrator employs a step size checking procedure, where the error of each step is estimated and if this exceeds a predefined error parameter the step size is reduced until the error is within allowed limits. It also increases the step size if the calculated error is below a certain limit, so the integration proceeds sufficiently quickly. The error parameter allows one to control the required accuracy of the integrator and plays an important role in ensuring that the orbits conserve energy.


\section{Ellipsoidal potential}\label{sec:ellipsoid}


To test that the algorithm produced the correct actions an output scheme was designed whereby the ``perfect ellipsoid" potential was used. This has actions which can be expressed analytically. The principle of the method involves fixing the actions, in a given potential, which thus define the initial conditions. On converting these to Cartesian coordinates one can then integrate the orbit in the ellipsoidal potential to get a set of Cartesian phase space points for the orbit and use the algorithm developed to extract the actions from this data. The extracted actions are then compared to the initial ones to measure the accuracy of the procedure.

All the details of the ellipsoidal potential are covered extensively in the classic paper by de Zeeuw \cite{deZeeuw}. In the first two parts of this section the results from this paper that are most relevant to this project are included.

\subsection{The ellipsoidal potential}
The ellipsoidal potential is generated from the density distribution
\begin{equation}
\rho = \frac{\rho_0}{(1+\tilde{m}^2)^2},
\end{equation}
where
\begin{equation}
\tilde{m}^2 = \frac{x^2}{a^2}+\frac{y^2}{b^2}+\frac{z^2}{c^2}, \quad a \geq b \geq c \geq 0.
\end{equation}
The potential is given in appendix \ref{sec:EllipsePot} and the parameters \(a, b\) and \(c\) provide the scale length along each axis of the ellipsoid.

There are four key types of orbit in the ellipsoidal potential (see figure 8 of \cite{deZeeuw}). Three of the types de Zeeuw calls ``tube" orbits, of which two lie along the x-axis and one along the z-axis; these have a well defined sense of rotation about these axes and thus one would expect to associate these with loop orbits in the isochrone. The other type of orbit is a box, and corresponds to that of an SHO in the limit of small oscillations around the centre of the potential. However, this box is not bound by flat planes as in the SHO case but by surfaces of hyperboloids and ellipsoids.

\subsection{Ellipsoidal coordinates}
In his paper de Zeeuw shows that the Hamilton-Jacobi equation is separable in ellipsoidal coordinates. Consequently the actions can then be expressed analytically, in a similar manner to the SHO case in \textsection\ref{sec:conversion}. Ellipsoidal coordinates are defined as the roots of the cubic equation for \(\tau\)
\begin{equation}\label{defEllipsoid}
\frac{x^2}{\tau-a^2} + \frac{y^2}{\tau-b^2} + \frac{z^2}{\tau-c^2} = 1.
\end{equation}
The three roots are labelled by \((\lambda, \mu, \nu)\) and they satisfy
\begin{equation}\label{restriction}
c^2 \leq \nu \leq b^2 \leq \mu \leq a^2 \leq \lambda.
\end{equation}
Surfaces of constant \(\lambda\) label ellipsoids, whilst surfaces of constant \(\mu\) and \(\nu\) label hyperboloids. Using equation \ref{defEllipsoid} one can write the ellipsoidal potential in ellipsoidal coordinates, this form can be found in appendix \ref{sec:EllipseCoordsPot}.


The solution of the Hamilton-Jacobi equation gives the momenta in ellipsoidal coordinates as
\begin{eqnarray}\label{ptau2}
p_\tau^2 = &A(\tau)& [( a^2 - \tau)(c^2 - \tau)E  \nonumber \\
&\quad&+( c^2 -\tau)i_2 \nonumber \\
&\quad&+( a^2 - \tau)i_3 + F(\tau)],
\end{eqnarray}
where \(E\) is the energy of the system, \(i_2, i_3\) are separation constants and
\[A(\tau) = [2(\tau-a^2)(\tau-b^2)(\tau-c^2)]^{-1}.\] The form of the known function \(F(\tau)\) can be found in appendix \ref{sec:EllipseFtau}. Note that the momenta are purely functions of one of the ellipsoidal coordinates only, reminiscent of the SHO case.

\subsection{Determining the range of \(\tau\)}\label{sec:rangeoftau}
It is important that the right hand side of (\ref{ptau2}) is positive, such that the momenta are real. As such, it is not sufficient that the \(\tau\) only satisfy the inequalities (\ref{restriction}), especially seeing as they do not put an upper bound on \(\lambda\). The range of positive values of each \(p_\tau^2\) is set by the parameters \(E, i_2, i_3, a, b\) and \(c\). The potential is fixed by choosing \(a, b\) and \(c\), then \(E, i_2\) and \(i_3\) are chosen such that the orbit is bound. For a bound orbit the total energy is negative, thus requiring \(E<0\) and that \(i_3 > 0\) \cite{deZeeuw}.

An algorithm was written to identify the range over which \(p_\tau^2\) is positive. This first identified the lower bounds by starting at the given lower bounds (\ref{restriction}) and incrementing \(\tau\) until \(p_\tau^2>0\). The value was then refined by stepping back once, decreasing the step size and incrementing until \(p_\tau^2>0\) again; repeating until a pre-defined accuracy was reached.

An upper bound on the \(\lambda\) term was estimated by incrementing from the lower bound up until \(p_\lambda^2\) became negative. This could be a number of orders of magnitude larger than the lower bound and thus a check was built in to increase the step size if \(p_\lambda^2\) was not decreasing fast enough. The upper bound was then found following a similar method as for the lower bound for all \(\tau\)s but approaching from above rather than below. The method also checks to ensure that the increment does not overshoot the limits imposed by (\ref{restriction}).

The values of the \(\tau\)s chosen for the initial conditions were taken to be the mean of the upper and lower boundary values for each \(\tau\). 

\subsection{Finding the initial conditions in Cartesian coordinates}
In order to initialise the orbit integrator the initial conditions need to be given in Cartesian coordinates. The transformation from ellipsoidal position coordinates to Cartesian position coordinates comes from the definition (\ref{defEllipsoid}) and is given in appendix \ref{sec:Ellipse2Cart}. In order to transform the momenta into Cartesian form the generating function \cite{PointTransform}
\begin{equation}
S(\vec{x}, \vec{p_\tau}) =  \vec{\tau}(\vec{x})\cdot \vec{p_\tau}
\end{equation}
is used. The form of the generating function arises because the ellipsoidal momenta are functions of one coordinate only. This then gives
\begin{equation}\label{px}
p_i =  \frac{\partial{\vec{\tau}}}{\partial{x_i}}\cdot \vec{p_\tau}.
\end{equation}

The partial derivatives in (\ref{px}) are found by inverting the Jacobian constructed from the \(\partial{x_i}/\partial{\tau}\) that are easily computed from the equations in appendix \ref{sec:Ellipse2Cart}. The inverted Jacobian is given in \ref{sec:EllipseJacobian}. 

The transformation made above for the momenta has been checked by computing the energy in both coordinate systems. The expression for the energy in ellipsoidal coordinates can be found in appendix \ref{sec:EllipseEnergy}. 

\subsection{Extracting the initial actions}

The three actions for the ellipsoidal potential correspond to one for each of the three coordinates \(\lambda, \mu, \nu\). The initial actions are calculated using a factor of four times equation \ref{actionInt}, where the range of the integral is over the regions of the \(\tau\)s that were found in \textsection \ref{sec:rangeoftau}. The factor of four arises in order that complete oscillations in each of the coordinates are considered \cite{deZeeuw}.

The function \(p_\tau(\tau)\) can be very steep as the boundary values of \(\tau\) are approached, so a suitable coordinate transformation is made to correctly sample these regions. This takes the form
\begin{eqnarray}
\tau  = &\bar{\tau} + \tau_{\Delta}\mathrm{sin}\vartheta, \nonumber \\
\mathrm{with} \quad\bar{\tau} = &\frac{1}{2}(\tau_{high} + \tau_{low}) \nonumber \\ 
\mathrm{and}\quad \tau_{\Delta}  = &\frac{1}{2}(\tau_{high} - \tau_{low}).
\end{eqnarray}
 Where \(\tau_{high}\) is the upper bound on the given \(\tau\), \(\tau_{low}\) is the lower bound on that \(\tau\) and \(\vartheta\) ranges between \(-\frac{\pi}{2} \leq \vartheta < \frac{\pi}{2}\).

\section{Results}\label{sec:results}


\begin{center}

\begin{figurehere}
\includegraphics[width=\linewidth]{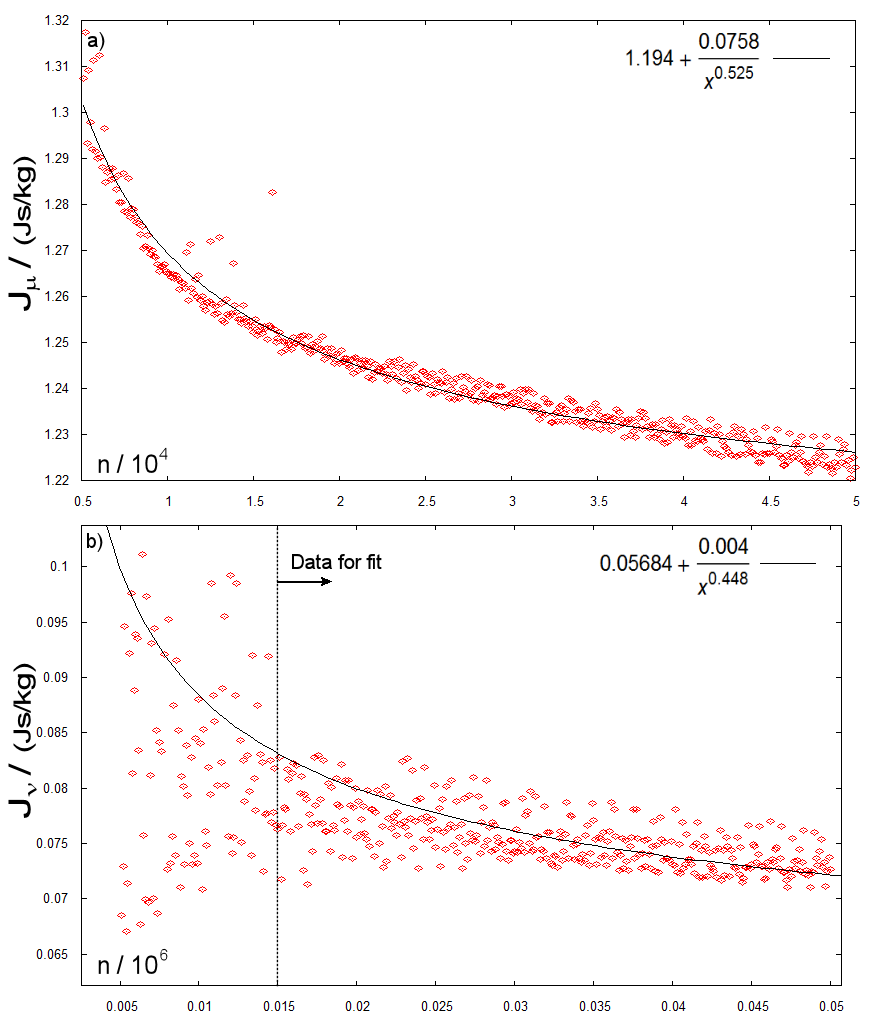}
\caption{\emph{Plots for the box orbit of \(J_\mu\) and \(J_\nu\) against \(n\), illustrating the smallest and largest discrepencies, respectively, between the computed and target actions.}}
\label{fig:ellipseVaryN}
\end{figurehere}
\end{center}

The direct output of the actions from the algorithm suggested that there was a significant deviation from the initial chosen actions. The reason for this forms the main part of the discussion in \textsection\ref{sec:discussion}. Essentially, it was found that this was because of insufficient sampling of the angle space. Figure \ref{fig:ellipseVaryN} presents examples of the results plotted for an increasing number of data points \(n\). The curve
\begin{equation}\label{Fit}
J_\tau = \frac{A}{n^m} + c,
\end{equation}
 was then fitted to the data, where \(A, m\) and \(c\) are fitting parameters, with \(m>0\). In the limit as \(n \rightarrow \infty \Rightarrow J_\tau = c\), giving a value of the action for infinite sampling. The data presented in tables \ref{loopEll} and \ref{boxEll} give the results of this fitting procedure. The initial actions are given as the target values. Table \ref{loopEll} are results for a loop orbit of the short axis tube type and those of table \ref{boxEll} are for a box orbit. The parameters for the potential used were \((a, b, c) = (10.2, 5.45, 3.25)\).

\begin{center}
\begin{tablehere}
\caption{Results for the short axis tube orbit. Using values for \((E, i_2, i_3) = (-23, 300.35, 10.8)\). Units of the actions are an arbitrary scale of J\(\cdot\)s/kg.}\label{loopEll}
  \begin{tabular}{ |c| c | c | c |}
    \hline
         &  $J_\lambda$  &  $J_\mu$ & $J_\nu$ \\ \hline
  target         & 11.66 &  47.31  & 0.1567 \\  \hline
  output       & 11.78 &  48.74  & 0.2418 \\  \hline
|\% diff.|	& 1.0 & 3.0 & 54 \\
    	\hline
  \end{tabular}
\end{tablehere}
\end{center}

\begin{center}
\begin{tablehere}
\caption{Results for the box orbit; *indicates the fit does not converge within 100 iterations of fitting algorithm. Using values for \((E, i_2, i_3) = (-31.02, -1928.39, 2.1)\). Units of the actions are an arbitrary scale of J\(\cdot\)s/kg.}\label{boxEll}
  \begin{tabular}{| c | c | c | c |}
    \hline
         &  $J_\lambda$  & $J_\mu$  & $J_\nu$ \\ \hline
     target    &   50.16 &  1.203  & 0.03390 \\ \hline
output & 51.43 &  1.194  & 0.05684* \\ \hline
|\% diff.| & 2.5 & 0.75 & 68\\
    	\hline
  \end{tabular}
\end{tablehere}
\end{center}

All but one of the fits of the data were calculated from data points in the range \(5\times 10^3 \rightarrow 50\times 10^3\) with an interval of \(100\). The box orbit \(J_\nu\) fit was taken between \(15\times 10^3\) and \(50\times 10^3\) due to noise at low \(n\). Only the \(50\times 10^3\) orbit needed to be integrated because all orbits lower than this could be extracted from the data set. This is beneficial as such orbit integration regimes can take a significant amount of time.

The upper limit of \(50\times 10^3\) existed because of memory limitations. In some cases an upper limit of the number of allowed points may also be set by the requirement that no two points may be coincident, as the FiEstAS algorithm cannot then calculate the volume of these two points. 


The results presented above demonstrate that the technique works to, at best, 0.75\% accuracy. However, it is also clear that this accuracy is not evenly distributed across the actions. Figure \ref{fig:ellipseVaryN} suggests a reason for this in that the data points do not form a smooth curve and that this becomes less smooth for the smaller targeted action (figure \ref{fig:ellipseVaryN}b). This noise causes a large uncertainty in the fit that is used and thus contributes to the lower accuracy of the result.

The integrals to produce the target actions were checked by decreasing the step size used in the integration and finding that the outputted value did not vary within at least four significant figures. As this was significantly less than the variation in the calculated actions using the algorithm then these were assumed to be the correct values.

The time taken to compute three actions is \(\approx 25s\) for a sample of \(50\times 10^3\) data points. This is comparable to the time taken  of \(\approx 15s\) for a two dimensional system using best-fit methods \cite{McMillan}. However, in order to do the fit as \(n\rightarrow\infty\) the actions need to be calculated multiple times and so increases computation time to the order of an hour.


\section{Discussion}\label{sec:discussion}

In this section the discrepancies are discussed and the reasons for them elucidated. The main contributing factor is an insufficient sampling of angle space due to an inhomogeneous distribution of angles. It is demonstrated that this also leads to the observed noise and so the reduced accuracy of the results.

\subsection{Clustering in angle space}
\begin{center}
\begin{figurehere}
\includegraphics[width=\linewidth]{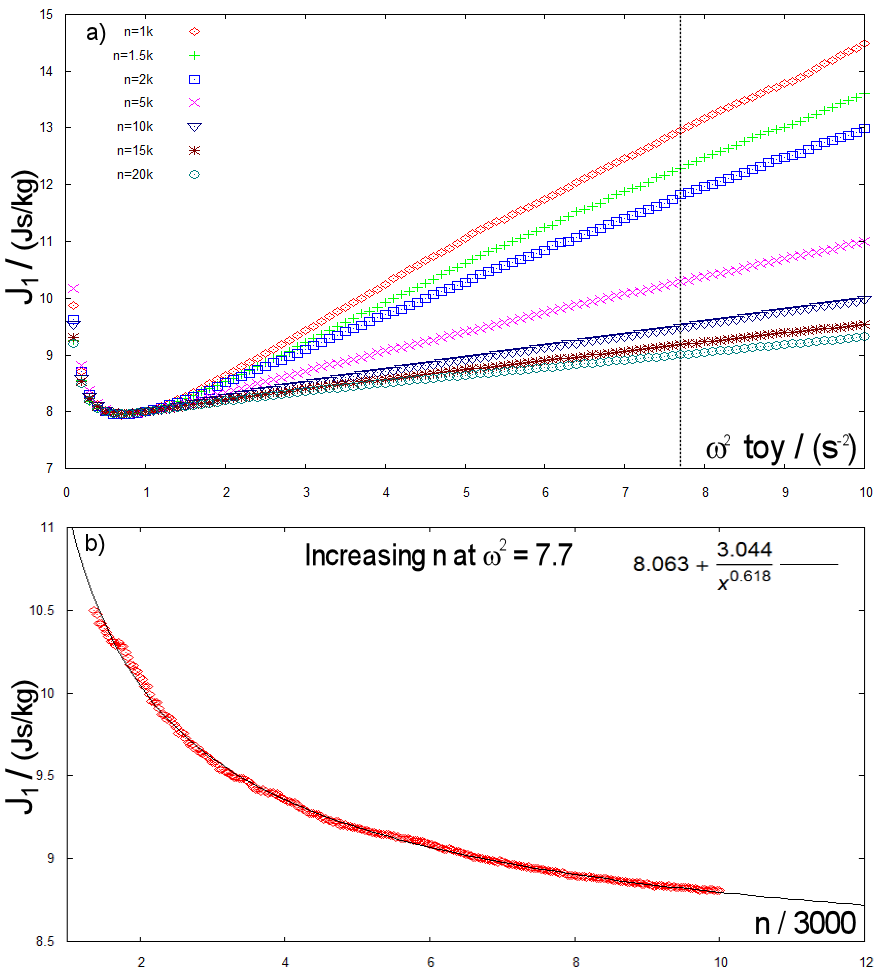}
\caption{\emph{a) Variation of outputted action \(J_1\) with increasing fitting parameter \(\omega_1^2\); with different curves for various \(n\). b) Selection of data points along the dotted line in a) at intervals of \(\Delta n = 100\), from \(n=4k\) to \(30k\).  }}
\label{fig:SHOactions}
\end{figurehere}
\end{center}
The main issue that reduces the accuracy of the method is a non-uniform distribution of points in angle space. In order to investigate this the simple case of orbits in a harmonic potential were considered. Choices of three frequencies, \(\omega_i^{\prime 2}\), were made for the target potential and the orbit integrated in this potential. A second set of frequencies, \(\omega_i^2\), were then chosen for the toy potential and the actions and angles corresponding to that toy potential calculated for the data points of the integrated orbit. The averaging procedure, as described in \textsection\ref{sec:algpract}, was then followed to calculate the actions.

The curves in figure \ref{fig:SHOactions}a display how the value of \(J_1\) calculated in the above method varies with the toy frequency. The value for \(\omega_1^{\prime 2} = 1\) was chosen, giving \(J'_1 = 8\). One can see that when the two frequencies match on the graph the action is correctly calculated by the algorithm. However, for increasing \(\omega_1^2\) the calculated action increases. The action increases also, after passing through a minimum, for decreasing \(\omega_1^2\). It is important to note that the target value of the action does not occur at the minimum of this plot.


The reason for the trend of increasing action with toy frequency appears to be because of a bias that is introduced into the averaging procedure by having the incorrect frequency. 
From (\ref{SHOtheta}) it is clear that if the frequency is higher than the target frequency, then the values of the angles assigned by the toy potential will cluster about \(0, \pi\) and \(2\pi\). This clustering corresponds to the apocentre of the orbit. There is then an increased density of points around apocentre so the volume associated with each point will be reduced. Considering equation \ref{SHOJ} for the actions, a larger value of the frequency at apocentre will increase the value of the action. The same result holds true for smaller frequencies, which cluster about pericentre.

As the averaging procedure sums over the product of the action at a data point and the corresponding volume, one would expect that the larger action would counteract the smaller volume and produce the correct target action. However, these two factors are not equal. At apocentre the action simply changes by a factor of \(\omega/\omega'\), whilst the angle has a similar reciprocal factor but non-linearity is introduced in taking the arctangent in (\ref{SHOtheta}). 

\begin{center}
\begin{figurehere}
\includegraphics[width=\linewidth]{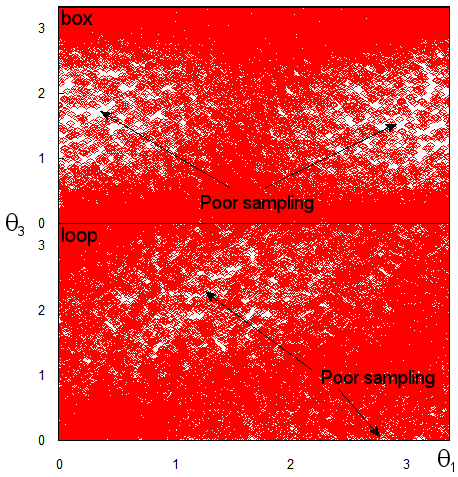}
\caption{\emph{Projection on to the plane \((\theta_1, \theta_3)\) showing the sampling of a subsection of angle space for n=50k, using the same initial conditions as the results in \textsection 8.}}
\label{fig:sampling}
\end{figurehere}
\end{center}

One would also expect to see such biasing in the isochrone case when the parameters for the fit do not match perfectly. This can be seen in figure \ref{fig:sampling}b, where even for a large sampling of angle space, there is clear inhomogeneity in the distribution of points. Figure \ref{fig:sampling} illustrates that this clustering occurs when both the SHO and isochrone potentials are used as toy potentials for orbits in the ellipsoidal potential. The bias is always going to occur in the ellipsoidal case because the potential in which the trajectory has been integrated will never match exactly either of the toy potentials.

It should also be noted that, even if the frequency matches, there are more points at apocentre than at pericentre because a star will naturally spend more time at apocentre than at pericentre due to it possessing a lower velocity at apocentre. This is somewhat corrected by the orbit integrator taking larger steps at apocentre, however, this does not generally compensate sufficiently. So there is always an underlying bias in the data set towards putting points at apocentre and this is emphasised further by the clustering seen for large frequencies.



\subsection{Insufficient sampling of angle space}\label{sec:insufsample}

In an attempt to increase the number density of points in the regions where the data had been skewed away from, the number of sampling points was increased. The different curves in figure \ref{fig:SHOactions}a show the effect of increasing the number of points, \(n\). It is clear that as one increases \(n\) the gradient of the curves decreases.

Plotting the values for a fixed toy frequency, taken at \(\omega^2 = 7.7\), against the number of sampling points one can see that the slope converges. Figure \ref{fig:SHOactions}b shows this data with a fit of equation \ref{Fit}. The value of the parameter \(c = 8.063\)  deviates from the target action of 8.0 by 0.8\%. The success of this method justified the use of the same technique in calculating the actions from the ellipsoidal potential. This approach should also work for low frequencies, though the divergence seen in figure \ref{fig:SHOactions}a as \(\omega \rightarrow 0\) may prove troublesome. Therefore, one would also require that \(\omega > 0\), which would only become an issue if the orbit being considered was confined to a plane. 

The reason that this extrapolation works is because taking the limit as \(n \rightarrow \infty\) in the discrete sum in (\ref{discreteAverage}) effectively reproduces the integral form of the equation that was introduced in (\ref{Jintegral}).

\subsection{Noise}
\begin{center}
\begin{figurehere}
\includegraphics[width=\linewidth]{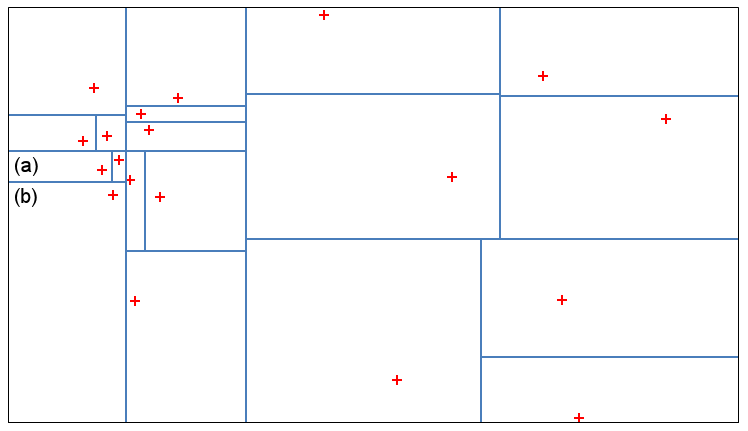}
\caption{\emph{Example of the FiEstAS algorithm on a two dimensional phase space with a cluster of points in the top left corner. Units are arbitrary.}}
\label{fig:Fiestas}
\end{figurehere}
\end{center}


The noise in figure \ref{fig:ellipseVaryN} becomes worse for smaller target actions. It is possible that this is because the angle space volumes are of order \(10^{-2}\), which compares to the same order of magnitude as the smallest target action. On this scale errors in the volume estimation by the FiEstAS algorithm become important.

The estimated angle space volume can be significantly different from the ``real" volume because in a space that has clusters of data points, the points on the boundary of the cluster can end up being assigned larger than expected volumes at the detriment of points in the sparser region. This can clearly be seen in figure \ref{fig:Fiestas}, which is a two dimensional example that illustrates the results FiEstAS produces. The small number of points in the figure were chosen to make it clear how the division process works and emphasise the problem of dealing with clusters. The volumes assigned to the points in (a) and (b) are significantly different, even though they are in regions of a similar density of points. This misallocation of volumes occurs because in order to be metric free FiEstAS only compares points along one coordinate axis at a time, which loses information on the density of points in the other coordinate directions.






\section{Conclusions}\label{sec:conclusions}

In this report action-angle coordinates have been introduced along with the key concept of a generating function. The theory of calculating actions via averaging procedures was presented along with an algorithm to complete the process numerically. The ellipsoidal potential and ellipsoidal coordinates were introduced and the principle of the method was demonstrated to work, achieving for some actions an accuracy of order \(1\%\) or better. The reason for the limited success was discussed and the sampling problem identified. A method to improve the results was developed and implemented, however, noise from the volume estimation hindered the effectiveness of this approach for small actions.

The averaging approach presented for calculating actions is currently restricted to orbits that can be fitted reasonably well by either an isochrone or SHO potential, such that the parameters allow for a smooth curve to be constructed to take the limit as \(n\rightarrow\infty\). The fit of the potential has to be reasonably good so that all the angles are not clumped together, causing FiEstAS to fail for high \(n\) integrations, which are required to construct an accurate infinite sampling fit.

To deal with orbits that are not fit well by either an isochrone or SHO potential it would be interesting to produce a code that would fit the ellipsoidal potential to an orbit. This would be beneficial because it could allow for a better covering of the transition region between a loop and a box orbit.

The issue of noise could be addressed by modifying the orbit integrator to increase the number density of points in the regions of angle space which are sparsely populated. This would then remove the problem of dealing with the boundaries between high and low density regions and give better estimates for the volumes. It could also reduce the insufficient sampling of angle space problem and thus the necessity to take the limit as \(n\rightarrow\infty\). However, there is some difficulty in identifying the required step size in Cartesian space that would produce a near uniform distribution of points in angle space. This is especially true in the isochrone case due to having to invert the complicated relationships between the two coordinate systems (appendix \ref{sec:AAana}). It would also mean that the orbit integration would have to be carried out a second time.

An alternative approach to dealing with the noise would be to use a metric based volume calculation\footnote{See the introduction of \cite{Ascasiber} for a brief review.}. The metric in this case would be trivial because it is only needed for the angle space. However, the disadvantage would then be the loss of the speed associated with the FiEstAS method. Thus, for a more accurate calculation it seems inevitable that a slower procedure is required.

Therefore, in comparison with the current best-fit methods \cite{McMillan} the method presented here still has a long way to be developed in order to compete both on computation time and accuracy. It may become useful so long as the problems discussed above could be resolved and then actions could be calculated accurately enough such that the \(n\rightarrow\infty\) fit would no longer be required.



\section{Acknowledgments}
I would like to thank my supervisor James Binney, as well as Paul McMillan, for their most useful guidance and support during this project.\raggedcolumns
\pagebreak
\end{multicols}

\pagebreak

\appendix



\section{Action-angle coordinates for the isochrone potential}\label{sec:AAana}
Here the relationship between Cartesian and action-angle coordinates is described for the case of the isochrone potential. This is taken from the derivation in \textsection 3.5.2 of \emph{Galactic Dynamics} \cite{BT}. The code to implement this conversion was provided by Paul McMillan.

For the isochrone case the actions are
\begin{eqnarray}
J_1 &=& L_z,\\
J_2 &=& L,\\
J_3 &=& \frac{GM}{\sqrt{-2E}} - \frac{1}{2} (L + \frac{1}{2}\sqrt{L^2 - 4GMb}).
\end{eqnarray}
Where \(E<0\) and is the total energy of the system, \(L\) is the total angular momentum, \(L_z\) is the angular momentum around the z-axis, being the axis of rotation, \(GM\) and \(b\) are the parameters of the potential. The angles have a slightly more complicated form, using spherical polar coordinates \((r,\vartheta,\phi)\) the first is given by
\begin{equation}
\theta_1 = \phi + \mathrm{sgn}(J_1)\int_{\pi/2}^{\vartheta}\frac{\mathrm{d}\vartheta}{\mathrm{sin}\vartheta\sqrt{\mathrm{sin}^2\vartheta\mathrm{sec}^2i - 1}}
\end{equation}
with \(i = \mathrm{arccos}(J_1/J_2)\). For the next two angles the following definitions are required
\begin{eqnarray}
s &=& 2 + \frac{c}{b}(1-e\mathrm{cos}\eta),\\
c &\equiv &\frac{GM}{-2E} - b,\\
e^2 &\equiv& 1 - \frac{J_2^2}{GMc}(1 + b/c),\\
s &\equiv& 1 + \sqrt{1+r^2/b^2},
\end{eqnarray}
where the relation to spherical polars comes in the last term with \(r^2\). So then the angles are
\begin{equation}
\theta_3 = \eta - \frac{ec}{c+b}\mathrm{sin}\eta
\end{equation}
and
\begin{eqnarray}
\theta_2 = \psi + \frac{1}{2}\left[1+ \frac{J_2}{\sqrt{J_2^2 + 4GMb}}\right]\left[\theta_3 - 2\mathrm{arctan}\left(\sqrt{\frac{1+e+2b/c}{1-e+2b/c}}\mathrm{tan}(\frac{1}{2}\eta)\right)\right]\nonumber \\
 - \mathrm{arctan}\left[\sqrt{\frac{1+e}{1-e}}\mathrm{tan}(\frac{1}{2}\eta)\right].
\end{eqnarray}
Where \(\mathrm{sin}\psi = \mathrm{cos}\vartheta/\mathrm{sin}i\).

\pagebreak

\section{Flow diagram of algorithm}\label{sec:FlowDiagram}

\begin{center}
\begin{figurehere}
\includegraphics[width=\linewidth]{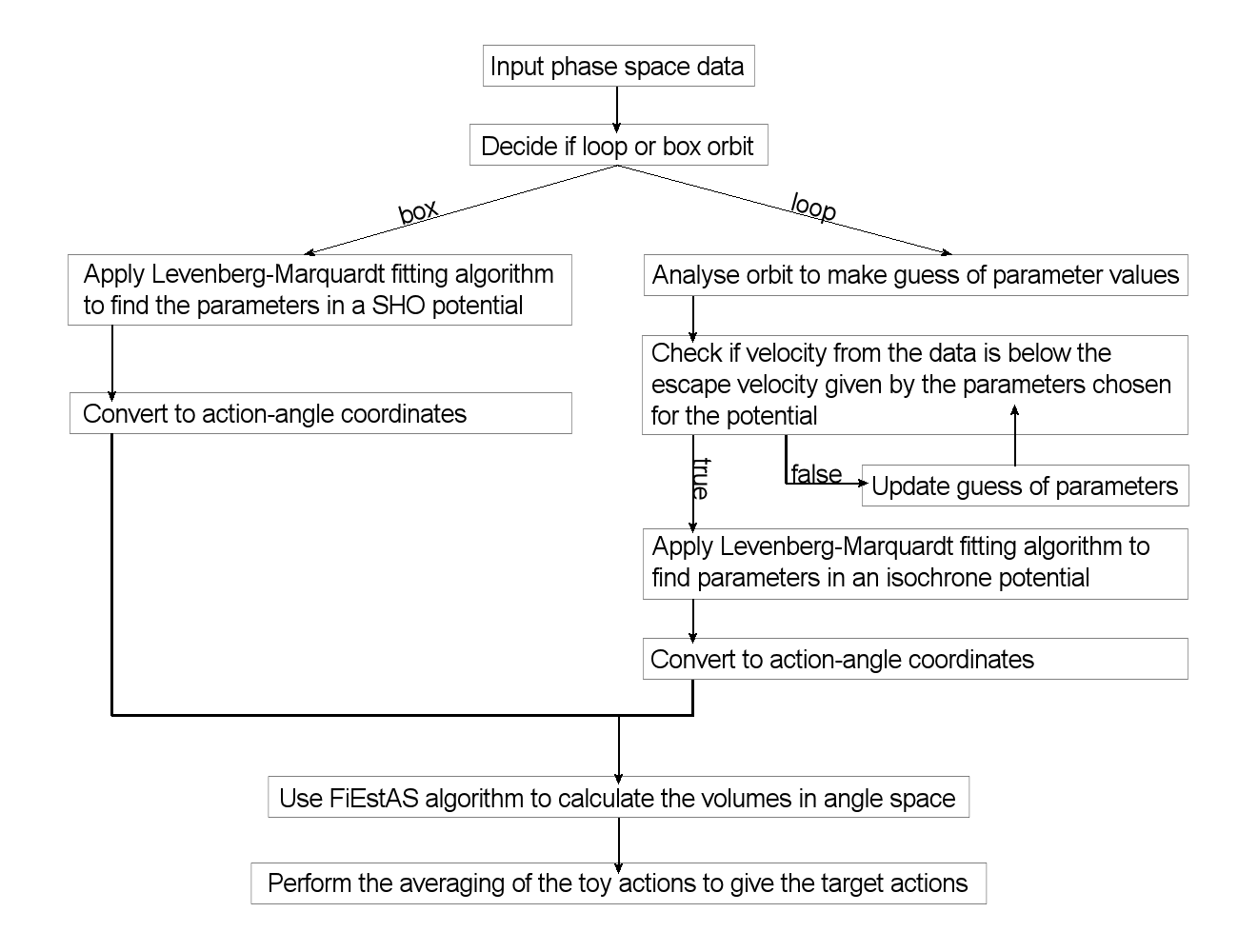}
\caption{\emph{Flow diagram of the algorithm used to calculate the actions given the Cartesian phase space coordinates of an orbit in an arbitrary potential.}}
\label{fig:FlowDiagram}
\end{figurehere}
\end{center}
\pagebreak

\section{Ellipsoidal coordinates}\label{sec:ellipsApp}
This appendix summarises the results from \cite{deZeeuw} which are relevant to this project, as well as the inverted Jacobian in \ref{sec:EllipseJacobian} that has been derived.

\subsection{Ellipsoidal potential in Cartesian coordinates}\label{sec:EllipsePot}
The ellipsoidal potential in Cartesian coordinates is,
\begin{equation}
V = - abc \int_0^\infty \frac{1}{(1+\frac{x^2}{(a^2+u)}+\frac{y^2}{(b^2+u)}+\frac{z^2}{(c^2+u)})}\frac{\mathrm{d}u}
{\sqrt{(a^2+u)(b^2+u)(c^2+u)}}
\end{equation}
where \(a, b, c\) are constant real parameters and the physical prefactor has been set to unity. The spatial derivative of this potential is what is used to calculate the acceleration in the orbit integrator. The integral is computed numerically by making the transformation
\[u=c^2 \mathrm{sinh}^2\vartheta\]
where the limits now run from \(0\rightarrow \vartheta_{max}\) where \(\vartheta_{max}\) is an upper limit chosen to replicate reaching infinity. Technically the upper limit of the integral in the new coordinate system is still infinity, however, the integrand dies off sufficiently quickly at large \(u\) and \(\vartheta\) that once \(u \gg a^2\) the integrand is practically zero. The choice of coordinate transformation simply allows for a quicker integration regime, as \(\mathrm{sinh}\vartheta\) goes to infinity much quicker than \(u\). 

\subsection{Ellipsoidal potential in Ellipsoidal coordinates}\label{sec:EllipseCoordsPot}

The ellipsoidal potential in ellipsoidal coordinates is
\begin{equation}\label{separableV}
V = - \frac{F(\lambda)}{(\lambda-\mu)(\lambda - \nu)}
- \frac{F(\mu)}{(\mu-\nu)(\mu-\lambda)} 
- \frac{F(\nu)}{(\nu-\lambda)(\nu-\mu)},
\end{equation}
with \(F(\tau)\) given in appendix \ref{sec:EllipseFtau} and \(\tau = \lambda, \mu, \nu\).

\subsection{The function \(F(\tau)\)}\label{sec:EllipseFtau}
With the physical parameters set to unity the function \(F(\tau)\) takes the form
\begin{equation}
F(\tau) = (\tau - a^2)(\tau -c^2)abc\int_0^\infty\frac{\sqrt{u+b^2}}{\sqrt{(u+a^2)(u+c^2)}}\frac{\mathrm{d}u}{u+\tau}.
\end{equation}
The integral is completed by numerical integration after transforming to coordinates \(u+c^2 = c^2/s^2\) and the range of integration runs between 1 and 0.

\subsection{Ellipsoidal to Cartesian coordinates}\label{sec:Ellipse2Cart}
The Cartesian position coordinates are given in terms of the ellipsoidal coordinates as
\begin{eqnarray}\label{cartesianfromellipse}
x^2 &=& \frac{(\lambda-a^2)(\mu-a^2)(\nu-a^2)}{( b^2-a^2)(c^2-a^2)},\nonumber\\
y^2 &=& \frac{(\lambda-b^2)(\mu-b^2)(\nu-b^2)}{(a^2-b^2)(c^2-b^2)},\nonumber\\
z^2 &=& \frac{(\lambda-c^2)(\mu-c^2)(\nu-c^2)}{(a^2-c^2)(b^2-c^2 )}.
\end{eqnarray}

\subsection{Jacobian}\label{sec:EllipseJacobian}
This is derived by taking the partial derivatives of the above relations (\ref{cartesianfromellipse}), constructing the Jacobian matrix and then inverting it to give 

\begin{equation}
\left(
\begin{array}{ccc}
\frac{\partial\lambda}{\partial x} &\frac{\partial\lambda}{\partial y} &\frac{\partial\lambda}{\partial z} \\
\frac{\partial\mu}{\partial x} &\frac{\partial\mu}{\partial y} &\frac{\partial\mu}{\partial z} \\
\frac{\partial\nu}{\partial x} &\frac{\partial\nu}{\partial y} &\frac{\partial\nu}{\partial z} \\
\end{array}
\right)
=
\left(
\begin{array}{ccc}
 \frac{2 x \left(\lambda -b^2\right) \left(\lambda -c^2\right)}{(\lambda -\mu ) (\lambda -\nu )} & \frac{2 y \left(\lambda -a^2\right) \left(\lambda -c^2\right)}{(\lambda -\mu ) (\lambda -\nu )} & \frac{2 z \left(\lambda -a^2\right) \left(\lambda -b^2\right)}{(\lambda -\mu ) (\lambda -\nu )} \\
 \frac{2 x \left(\mu -b^2\right) \left(\mu -c^2\right)}{(\mu -\lambda ) (\mu -\nu )} & \frac{2 y \left(\mu -a^2\right) \left(\mu -c^2\right)}{(\mu - \lambda) (\mu -\nu )} & \frac{2 z \left(\mu -a^2\right) \left(\mu -b^2\right)}{( \mu-\lambda ) (\mu -\nu )} \\
 \frac{2 x \left(\nu -b^2\right) \left(\nu -c^2\right)}{(\nu-\lambda  ) (\nu -\mu )} & \frac{2 y \left(\nu -a^2\right) \left(\nu -c^2\right)}{(\nu - \lambda) (\nu -\mu )} & \frac{2 z \left(\nu -a^2\right) \left(\nu -b^2\right)}{(\nu -\lambda ) (\nu -\mu )} \\
\end{array}
\right),
\end{equation}
and with \(x, y, z\) replaced by the expressions in \ref{sec:Ellipse2Cart}.

\subsection{Energy in ellipsoidal coordinates}\label{sec:EllipseEnergy}
The energy is given by
\begin{equation}
H = \frac{p_\lambda^2}{2P^2} + \frac{p_\mu^2}{2Q^2} + \frac{p_\nu^2}{2R^2} + V(\lambda, \mu, \nu),
\end{equation}
where the \(P, Q, R\) are the metric coefficients
\begin{eqnarray}
P^2 &=& \frac{(\lambda - \mu)(\lambda-\nu)}{4(\lambda-a^2)(\lambda-b^2)(\lambda-c^2)},\nonumber\\
Q^2 &=& \frac{(\mu-\nu)(\mu-\lambda)}{4(\mu-a^2)(\mu-b^2)(\mu-c^2)},\nonumber\\
R^2 &=& \frac{(\nu - \lambda)(\nu-\mu)}{4(\nu-a^2)(\nu-b^2)(\nu-c^2)}.
\end{eqnarray}

\end{document}